\documentclass{aa}

\usepackage{graphicx}
\usepackage{txfonts}
\usepackage{graphicx}    
\usepackage{amsmath}    
\usepackage{float} 
\usepackage{graphicx} 
\usepackage{booktabs} 
\usepackage{hyperref} 
\usepackage{geometry}
\usepackage{ifthen}
\usepackage{lipsum}
\usepackage{color, soul} 
\usepackage{makecell}
\usepackage{subcaption}
\usepackage{xspace} 
\usepackage{multirow}
\usepackage{tablefootnote}

\usepackage[normalem]{ulem}
\usepackage[dvipsnames]{xcolor}

\definecolor{seagreen}{rgb}{0.190, 0.525, 0.361}

\newcommand{\eq}[2][]{\begin{align}
                          #2
\end{align}}

\newcommand{\abs}[1]{\left | #1 \right |}
\newcommand{\avg}[1]{\left \langle #1 \right \rangle}

\newcommand{\lr}[1]{\left(#1\right)}

\newcommand{\BH}{\mathrm{BH}}

\newcommand{\units}[1]{\,\mathrm{#1}}

\newcommand{\Gaia}{{\it Gaia}\xspace}
\newcommand{\gb}{\Gaia~BH3\xspace}
\newcommand{\feh}{\mathrm{[Fe/H]}}
\newcommand{\sbh}{S-BH\xspace}
\newcommand{\sbhs}{S-BHs\xspace}

\newcommand{\thetaG}{\mathrm{G_3}\xspace}

\def\subinrm#1{\sb{\rm#1}}
    {\catcode`\_=13 \global\let_=\subinrm}
\mathcode`\_="8000
\def\upsubscripts{\catcode`\_=12 } 

\upsubscripts

\newcommand{\msun}{{\rm M}_\odot}

\renewcommand{\eqref}[1]{(equation~\ref{#1})}

\begin{document}

\title{$N$-body modelling of the ED-2 stream progenitor shows \\\textit{Gaia} BH3's formation involved dynamical interactions}
\titlerunning{$N$-body modelling of the ED-2 progenitor shows \textit{Gaia} BH3's formation involved dynamics} 

\author{Daniel Marín Pina
\inst{1,2,3}
\and
Mark Gieles
\inst{2,4,5}
\and
Sara Rastello 
\inst{2,3}
\and
Clàudia Garcia-Diago
\inst{2,3}
\and
Giuliano Iorio
\inst{2,3}
\and
Marc Ardèvol
\inst{2,3}
}

\institute{
    Zentrum für Astronomie der Universität Heidelberg, Institut für Theoretische Astrophysik, Albert-Ueberle-Str. 2, 69120 Heidelberg\\    \email{daniel.marin@uni-heidelberg.de}
    \and
    Institut de Ciències del Cosmos (ICCUB), Universitat de Barcelona (UB), c. Martí i Franqués, 1, 08028 Barcelona, Spain
    \and
    Departament de F\'isica Qu\`antica i Astrof\'isica (FQA), Universitat de Barcelona (UB), Mart\'i i Franqu\`es 1, 08028 Barcelona, Spain
    \and
    ICREA, Pg. Llu\'is Companys 23, 08010 Barcelona, Spain
    \and
    Institut d'Estudis Espacials de Catalunya (IEEC), Edifici RDIT, Campus UPC, 08860 Castelldefels (Barcelona), Spain
}

\date{Received XX; accepted XX}


\abstract
{The \textit{Gaia} collaboration announced the discovery of a binary of a massive black hole ($33\,\msun$) with a low-mass giant star (\textit{Gaia} BH3) in the ED-2 stellar stream. The properties of this binary, as well as its position in the stream, challenge a formation scenario invoking only isolated binary evolution.}
{We aim to quantify the importance of cluster dynamics in the formation of \gb in the progenitor cluster of the ED-2 stream.} 
{We perform detailed $N$-body simulations of the progenitor cluster of the ED-2 stream, including the effects of single and binary stellar evolution. We compare these simulations to observations of the ED-2 stream and the properties of \gb.}
{We determine that \gb most likely formed as an exchange binary which underwent multiple strong dynamical interactions. We highlight the importance of cluster dynamics in assembling \gb, and disfavour a formation scenario where it evolved unperturbed by dynamical interactions.}
{The role of dynamics should be considered when interpreting the properties of the population of star-black hole binaries found in the next \textit{Gaia} Data Release.}


\maketitle
%

\section{Introduction}

Traditionally, observational evidence of (stellar-origin) black holes (BHs) came from X-ray binaries \citep{CorralSantana2015}. This method is biased towards detecting binaries that are sufficiently tight to allow mass transfer from the star to the BH. In the last decade, most of the known BHs have been detected through gravitational wave (GW) emission \citep{TheLIGOScientificCollaboration2025}. These are produced in the last fractions of a second before a binary BH (BBH) merges, so the observational bias towards close binaries is even more dramatic.

Recently, we have begun to uncover a population of  BHs that do not merge or accrete; they are `dormant' BHs. One of the methods to detect dormant BHs is in star-BH (S-BH) binaries via astrometry, measuring the perturbation of the parallax and proper motion of the companion star. In this regard, the \textit{Gaia} mission has been key to uncovering these systems, with confirmed detections of three dormant BHs, named \textit{Gaia} BH1 \citep{elbadry_bh1, Chakrabarti2023}, \textit{Gaia} BH2 \citep{el-badry_bh2}, and \textit{Gaia} BH3 \citep{GaiaBH3}. Many more such systems are expected to be discovered in the upcoming fourth \textit{Gaia} Data Release (DR4). On top of these, other dormant S-BH binaries have been discovered with other methods, such as spectroscopy (VFTS 243, \citealt{Shenar2022}; NGC 3201\_12560, \citealt{Giesers2018, Giesers2019}; HD 130298, \citealt{Mahy2022}; NGC 1850 BH1, \citealt{Saracino2022, Saracino2023} (but see \citealt{ElBadry2022}); \textit{Gaia} DR3 3425577610762832384, not yet confirmed, \citealt{Wang2024}), lensing (OGLE-2011-BLG-0462, \citealt{Lam2022}), and most recently, Hubble/JWST astrometry \citep{Whitaker2026}\footnote{See \href{https://mkenne15.github.io/BHCAT/index.html}{mkenne15.github.io/BHCAT} for an updated list of dormant BHs.}.

The properties and orbits in the Galactic plane of \textit{Gaia} BH1 and \textit{Gaia} BH2 are consistent with formation in star clusters \citep{2023MNRAS.526..740R, 2024ApJ...965...22D, Tanikawa2024, Arcasedda2024-dragon2},  isolated stellar binaries \citep{Kotko2024, Mapelli2026}, or stellar triples \citep{Li2026}. Both \textit{Gaia} BH1 and \textit{Gaia} BH2 have relatively light BHs ($\lesssim 10\units{\msun}$), mildly metal-poor companions ($\feh \sim -0.2$), and they orbit in the Galactic plane. Compared to these, the properties and Galactic orbit of \gb stand out. This binary consists of a massive BH ($m_\BH = 33\units{\msun}$) and an extremely metal-poor ($\feh = -2.56$), low-mass ($m_\star=0.76\units{\msun}$) giant star. The binary has a long period ($P=4\times10^3\units{d}$) and has an eccentricity of $e=0.7$. \gb is remarkably close to the Sun ($590\units{pc}$), in an orbit in the Milky Way (MW) halo associated with the ED-2 stellar stream, which probably formed from a now-dissolved star cluster \citep{Balbinot2023}. The association with the stream is both kinematical and chemical \citep{Balbinot2024, Hackshaw2025, VandenBroeck2026}. The mass of the BH is much larger than any other (stellar-origin) BH detected in the MW, and is close to the median BH mass detected via GWs \citep[see Fig.~3 in][]{MarinPina2024}. This binary bridges the populations of BHs detected via electromagnetic measurements and GWs; hence, understanding its formation pathway can provide constraints on the populations of binaries with BHs and on the formation of GW sources.

There are two proposed explanations for the origin of \gb: either it formed from the isolated evolution of a binary of stars \citep[see e.g.][]{Iorio2024, ElBadry2024} or it formed via dynamical interactions in the progenitor cluster of the ED-2 stream \citep{MarinPina2024}. Both of these channels predict no chemical peculiarities of the \gb star with respect to other stars in ED-2, in line with recent chemical analyses \citep[e.g.,][]{Hackshaw2025, VandenBroeck2026}. The fact that \gb is associated with the ED-2 stream provides circumstantial evidence that dynamics played a role in shaping its present-day properties, but does not exclude a binary origin unperturbed by dynamics. In this paper, we model the collisional dynamics of the progenitor cluster of the ED-2 stream to quantify the effect of dynamics in the formation of \gb. A preliminary analysis of the results presented hereafter was shown in the conference proceedings \cite{MarinPina2025}.

In Sect.~\ref{sec:methods}, we describe the $N$-body models used to simulate the ED-2 stream progenitor. In Sect.~\ref{sec:stream}, we study the stream formation in our simulations and constrain the cluster's initial parameters with the observed \textit{Gaia} data. In Sect.~\ref{sec:gbh3}, we explore the most likely origin of \gb, including the hypothesis that it is a triple system with an inner BBH. In Sect.~\ref{sec:formenv}, we characterise the formation environment of S-BH binaries in our simulations. In Sect.~\ref{sec:discussion} and Sect.~\ref{sec:conclusions}, we discuss and summarise our findings.

\section{Methods}
\label{sec:methods}
We performed a set of simulations of the ED-2 progenitor cluster using the high-performance $N$-body code \textsc{petar} \citep{Wang2020}. \textsc{petar} uses a hybrid method to integrate the evolution of dense collisional systems, which separates the long- and short-range interactions. The former are evolved using a Barnes-Hut particle-tree method \citep{BarnesHut1986} with a second-order leapfrog integrator; the latter are integrated using a fourth-order Hermite method with slowdown-algorithmic regularisation \citep[SDAR,][]{SDAR}. The SDAR method uses a change of coordinates to avoid singularities in Newtonian gravity \citep[algorithmic regularisation,][]{KustaanheimoStiefel1965} together with a time scaling \citep[`slowdown',][]{MikkolaAarseth1996}, which is used to efficiently compute the perturbation of a binary over several orbits. This method allows us to avoid the computational cost and numerical instability caused by having very small timesteps. With this hybrid formalism (particle-tree and SDAR), we can efficiently simulate star clusters with a significant binary population.

The initial conditions for the cluster simulations are set up using the \textsc{mcluster} code \citep{Kupper2011}. We initialise all models using a \cite{Plummer1911} density profile with a \cite{Kroupa2001} initial mass function (IMF) between $0.08\units{\msun}$ and $150\units{\msun}$. The initial mass and half-mass density of the ED-2 progenitor, $M_0$ and $\rho_{h, 0}$, are not known. Our approach is to fix the (approximate) dissolution time of the cluster to $t_{dis}\in[0.5, 4, 8]\units{Gyr}$, for different combinations of $M_0$ and $\rho_{h,0}$. The method used to compute them is explained in Appendix~\ref{sec:appendix:diss}.  The range in masses of all models is within the constraints imposed by \cite{Balbinot2024}, namely $M_0 \gtrsim 2\times 10^3\units{\msun}$ (such that at least one massive star that can create a BH is sampled from the IMF) and $M_0 \lesssim 5.2\times 10^4\units{\msun}$ (because of the neglible scatter in light element abundances). 

\begin{table*}[h]
\caption{Initial parameters of the cluster simulations: model name, mass $M_0$, half-mass density $\rho_{h, 0}$, (approximate) dissolution time $t_{dis}$ (computed according to Appendix~\ref{sec:appendix:diss}), binary fraction $f_b$, and number of runs with different random seeds $N_{runs}$.}
\label{table:initial_params}
\centering 
\begin{tabular}{c c c c c c}
\hline\hline
Model name & $t_{dis}$ [Gyr] & $\rho_{h, 0}$ [$\mathrm{\msun}$ pc$^{-3}$] & $M_0$ [$\mathrm{\msun}$] & $f_b$ & $N_{runs}$\\  
\hline   
TD8\_HiRHO\_WBIN & \multirow{2}{*}{8} & \multirow{2}{*}{760} & \multirow{2}{*}{$7.7\times 10^3$} & 0.4 & 10 \\
TD8\_HiRHO\_NOBIN &                    &                      &                                   & 0.0 & 10 \\
TD8\_LoRHO\_WBIN & \multirow{2}{*}{8} & \multirow{2}{*}{240} & \multirow{2}{*}{$3.9\times 10^4$} & 0.4 & 10 \\
TD8\_LoRHO\_NOBIN &                    &                      &                                   & 0.0 & 10 \\
TD4\_HiRHO\_WBIN & \multirow{2}{*}{4} & \multirow{2}{*}{760} & \multirow{2}{*}{$2.7\times 10^3$} & 0.4 & 10 \\
TD4\_HiRHO\_NOBIN &                    &                      &                                   & 0.0 & 10 \\
TD4\_LoRHO\_WBIN & \multirow{2}{*}{4} & \multirow{2}{*}{240} & \multirow{2}{*}{$1.4\times 10^4$} & 0.4 & 10 \\
TD4\_LoRHO\_NOBIN &                    &                      &                                   & 0.0 & 10 \\
TD0.5\_VLoRHO\_WBIN &  0.5 & 25 & $3.5\times 10^3$ & 0.4 & 10 \\
\hline
\end{tabular}
\end{table*}

\begin{figure*}[htbp]
    \centering
    \includegraphics[width=17cm]{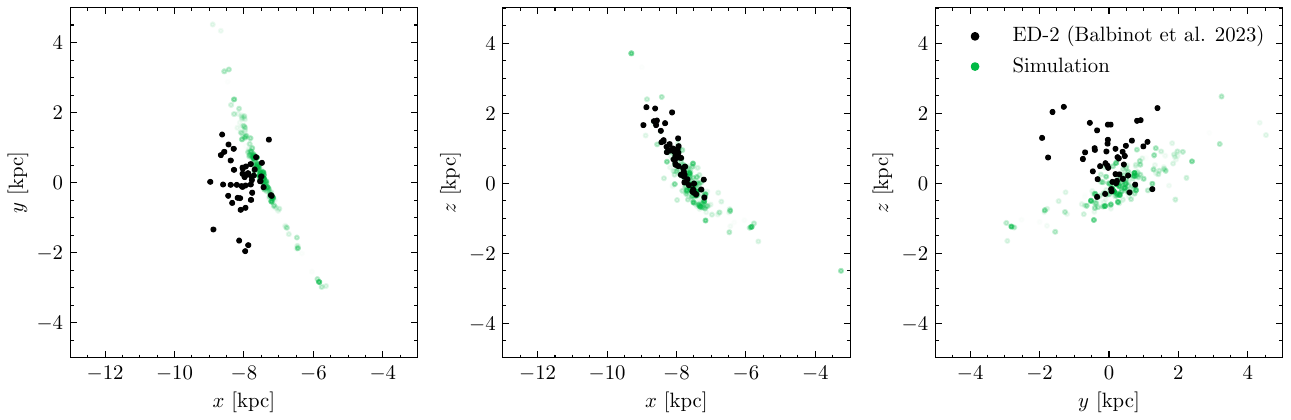}
    \caption{Positions of the stars of the ED-2 stream in Galactocentric cartesian coordinates. In black, \Gaia data from \cite{Balbinot2023}. In green, a run of the TD8\_LoRHO\_NOBIN model, with each star coloured according to its magnitude. The cluster is shown at the end of the simulation ($13\units{Gyr}$), which is set up such that the centre of mass of the cluster matches the observed current position of \gb.}
    \label{fig:stream}
\end{figure*}

These models are run both with and without a primordial binary population. Hereafter, we use the term `primordial' to refer to binaries that originated as part of the cluster formation process. For the models with primordial binaries, we use 100\% binary fraction among stars above $5\units{\msun}$, using the \cite{Sana2012} period and eccentricity distributions. We acknowledge that this is a mild extrapolation of primary masses with respect to their sample, as well as being on the higher end of initial binary fraction. For lower-mass stars, we use the \cite{DuquennoyMayor1991} period distribution and a thermal eccentricity distribution, with a total binary fraction of $f_b=0.4$. The low- and high-mass stars are paired in binaries separately, and thus we do not allow extreme-mass-ratio binaries at the beginning of the simulation (the minimum mass ratio at zero-age main sequence is 1/30, larger than that required to form \textit{Gaia} BH3). This choice, motivated by the data showing that O and B stars are more likely to be found in equal-mass binaries \citep{SanaEvans2011}, is discussed in Sect.~\ref{sec:discussion:bse}. We assume that the clusters have a single population of stars, with metallicity equal to that of the \gb star, $\feh = -2.56$ ($Z\simeq 6\times10^{-5}$). The initial parameters of the cluster models are summarised in Table~\ref{table:initial_params}. 

Once the models are set up, we evolve them using \textsc{petar} until $13\units{Gyr}$. Due to the low metallicity, we perform the single and binary stellar evolution with the \textsc{bseEmp} code \citep{Tanikawa2020, Tanikawa2022}. We assume that compact object formation follows the rapid SN mechanism of \cite{Fryer2012}, with a strong pulsational pair instability (PPSN) cutoff \citep{Belczynski2016}. Overall, our models have a range from a few to a hundred BHs.

In order to include the effect of the Milky Way tidal field on the clusters, \textsc{petar} can interface with \textsc{galpy} \citep{galpy}, a code library used to model Galactic dynamics. In our simulations, the clusters are evolved in a Milky Way-like potential using the \textsc{mwpotential2014} routine. The initial conditions are such that, after $13\units{Gyr}$, the centre of density of the cluster (modelling it as a test particle) matches the current position of \gb in the sky. In the following Section, we relax this assumption and consider shifts of an extra $\Delta t \leq 100\units{Myr}$ in the orbital evolution. The Galactic orbital elements are computed using \textsc{galpy} and the \textsc{petar} prescriptions for the Sun's position and velocity, and shown in Table~\ref{table:galaxy_orbit}. 

\begin{table}[h]
\caption{Initial Galactic orbit of the cluster at the start of the simulation, derived from backtracking the estimated Galactic orbit of \gb in the \textsc{mwpotential2014} potential, assuming it matches the position of the centre of mass of the cluster at $t=13\units{Gyr}$.}
\label{table:galaxy_orbit}
\centering 
\begin{tabular}{l c c}
\hline\hline
 & $t=13\units{Gyr}$ & $t=0$ \\  
\hline   
Galactocentric $x$ [kpc] & -7.63 & -0.222 \\
Galactocentric $y$ [kpc] & 0.466 & -8.23 \\
Galactocentric $z$ [kpc] & -0.022 & -2.98 \\
Galactocentric $v_{\it x}$ [km/s] & 121 & 266 \\
Galactocentric $v_{\it y}$ [km/s] & -280 & 164 \\
Galactocentric $v_{\it z}$ [km/s] & -134 & 19.3 \\
Orbital SMA $a$ [kpc] & \multicolumn{2}{c}{17.1} \\
Orbital eccentricity $e$ & \multicolumn{2}{c}{0.616} \\
Radial period $T_p$ [Myr] & \multicolumn{2}{c}{406} \\
\hline
\end{tabular}
\end{table}

\section{Formation of the ED-2 stream}
\label{sec:stream}
To verify that our simulations are representative of the progenitor cluster of the ED-2 stream, we compare the streams formed in our simulations to the ED-2 \Gaia data reported in \cite{Balbinot2023}. Since the cluster is old \citep[$t\gtrsim 13\units{Gyr}$,][]{Balbinot2024}, it must have undergone several orbits around the Milky Way. There are many uncertainties associated with the evolution of the Galactic potential, so our goal is not to match the shape of the observed stream exactly, but rather to show that our models dissolve and form a structure similar to ED-2.

Since the cluster is completely dissolved, we expect that most of the stars will have dispersed through the Galaxy and have very faint magnitudes. To account for detectability, we convert the output of our simulations (based on the radii and luminosities of the stars from \textsc{bseEmp}) to the \Gaia $G$ magnitudes. We first compute the bolometric magnitude, then use the $B-V$ colour correction from \cite{Flower1996, Torres2010} and the $G-V$ corrections from \cite{Jordi2010}. We then limit our sample to stars within the $G$ magnitude limits in the ED-2 sample from \cite{Balbinot2023}. For illustrative purposes, in Fig.~\ref{fig:stream} we show the Galactocentric positions of the members of ED-2 \citep{Balbinot2023} superimposed on the final snapshot of one of our simulations (TD8\_LoRHO\_NOBIN).

\begin{figure}
    \begin{subfigure}{\columnwidth}
    \resizebox{\hsize}{!}{\includegraphics{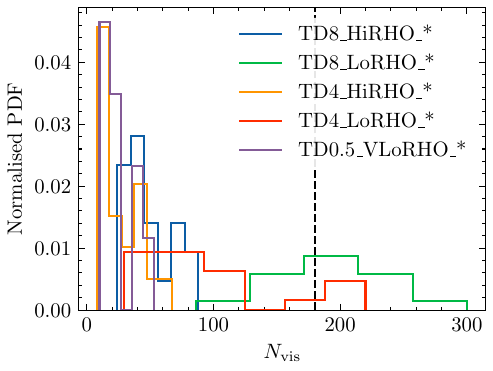}}
    \caption{Obtained at the end of the simulation ($13\units{Gyr}$)}
    \label{fig:Nvis_130}
    \end{subfigure}
    
    \bigskip
    
    \begin{subfigure}{\columnwidth}
    \resizebox{\hsize}{!}{\includegraphics{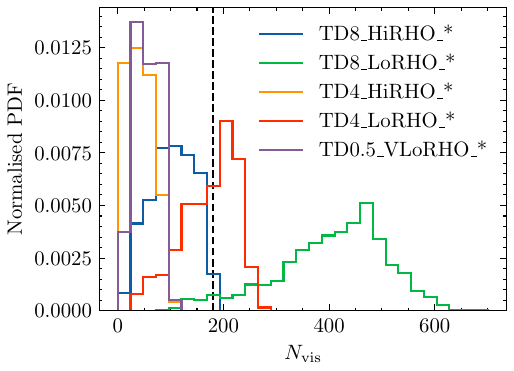}}
    \caption{Obtained from snapshots uniformly spaced in the $100\units{Myr}$ after the end of the simulation.}
    \label{fig:Nvis_evo}
    \end{subfigure}

    \caption{ \label{fig:Nvis_all}Probability distribution function (PDF) of the number of visible sources in the stream, separated by initial conditions of the simulations. The models with and without primordial binaries are binned. In dashed black, $N_{vis}=180$ \citep{Mikkola2023}.}

\end{figure}

Per \cite{Mikkola2023}, the ED-2 stream is expected to contain roughly $N_{vis}=180$ observable members within $3\units{kpc}$ from the Sun (which are not found because of the lack of radial velocity measurements). This estimate is obtained with a data-driven maximum penalised-likelihood algorithm, allowing the authors to obtain the number of stars in the stream while lacking full radial velocity measurements. In Fig.~\ref{fig:Nvis_130}, we compare their value for $N_{vis}$ to what is found in the final snapshot of each of our simulations. We preface our analysis of this comparison by stating that the $N_{vis}=180$ value is a rough estimate; as such, we use it as a `sanity-check' verification. Furthermore, we note that the stream formation process is stochastic and, as such, we do not aim for an exact match, just qualitative agreement. From this comparison, we see that the most massive models are preferred by the data. The low-density, early dissolution model (TD0.5\_VLoRHO\_*) results in too low $N_{vis}$, and thus we exclude it from further analysis.

\begin{figure*}[htbp]
    \centering
    \includegraphics[width=17cm]{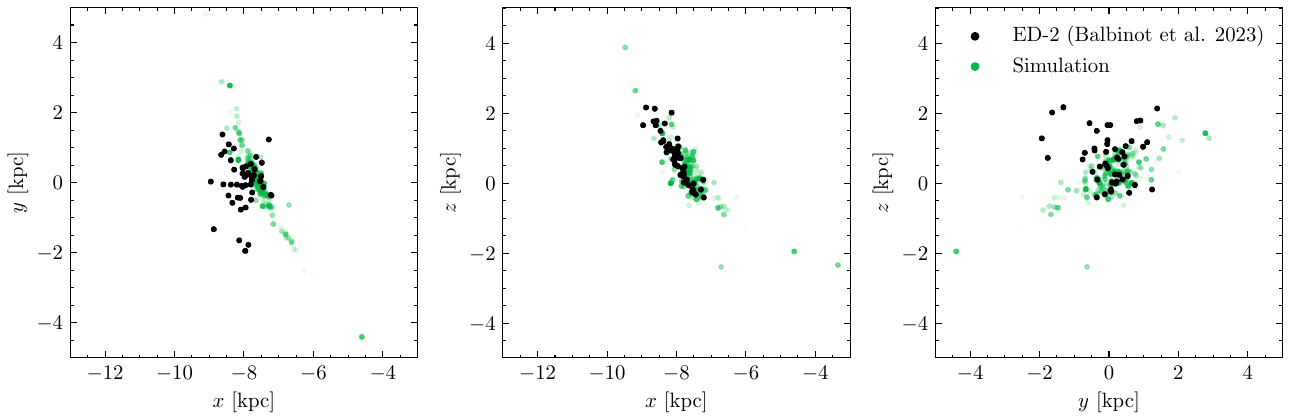}
    \caption{Positions of the stars of the ED-2 stream in Galactocentric cartesian coordinates. In black, \Gaia data from \cite{Balbinot2023}. In green, a run of the TD8\_HiRHO\_NOBIN model, with each star coloured according to its magnitude. The cluster is shown $37\units{Myr}$ after the end of the simulation, to maximise visual similarity to the ED-2 stream.}
    \label{fig:stream2}
\end{figure*}

In Fig.~\ref{fig:Nvis_130}, we assume that the observed Galactocentric position of \gb corresponds to the centre of the stream. While this is approximately true for the observed stars, it might not be the case for the complete stream. To ensure that our conclusions are independent of this choice, we evolve the stream for an additional $100\units{Myr}$ (about a quarter of the radial period), taking 10000 snapshots per model, equally spaced in time. For illustrative purposes, we show one of such snapshots in Fig.~\ref{fig:stream2} (from model TD8\_HiRHO\_NOBIN, after $37\units{Myr}$), chosen such that $N_{vis}$ is equal to the value found in \cite{Mikkola2023}. In Fig.~\ref{fig:Nvis_evo}, we show $N_{vis}$ for these snapshots, separated by the initial parameters. A similar conclusion can be drawn as for Fig.~\ref{fig:Nvis_130}: the more massive clusters are preferred over lower mass clusters.

\section{The origin of \gb}
\label{sec:gbh3}
In the previous Section, we show that our models of GCs dissolve in the \gb orbit and form streams similar to ED-2, therefore lending support to the hypothesis that \gb formed in the ED-2 progenitor cluster. In a cluster, we distinguish the following formation channels for \gb:
\begin{itemize}
    \item `primordial', where the binary originated as part of the cluster formation process, but interactions with other stars, BHs, and binaries altered its properties (period and eccentricity) significantly;
    \item `exchange', where the star or BH formed in a primordial binary but exchanged its companion via dynamical interactions;
    \item `dynamical', where the binary was assembled from a previously-unbound star and a BH (or its progenitor star) via dynamical interactions. 
    \item `isolated', where the binary originated as part of the cluster formation process and evolved unperturbed by dynamical interactions.
\end{itemize}
In Sect.~\ref{ssec:formation}, we explore the \sbh systems formed in the first three scenarios, and obtain the most likely formation channel for \gb in a dynamical environment. In Sect.~\ref{ssec:iso}, we explore the last scenario, quantifying whether dynamical interactions have a significant effect on the formation of \sbh systems in the ED-2 stream progenitor. 

\subsection{Formation of \gb in the ED-2 progenitor}
\label{ssec:formation}
\begin{figure*}[htbp]
    \centering
    \includegraphics[width=17cm]{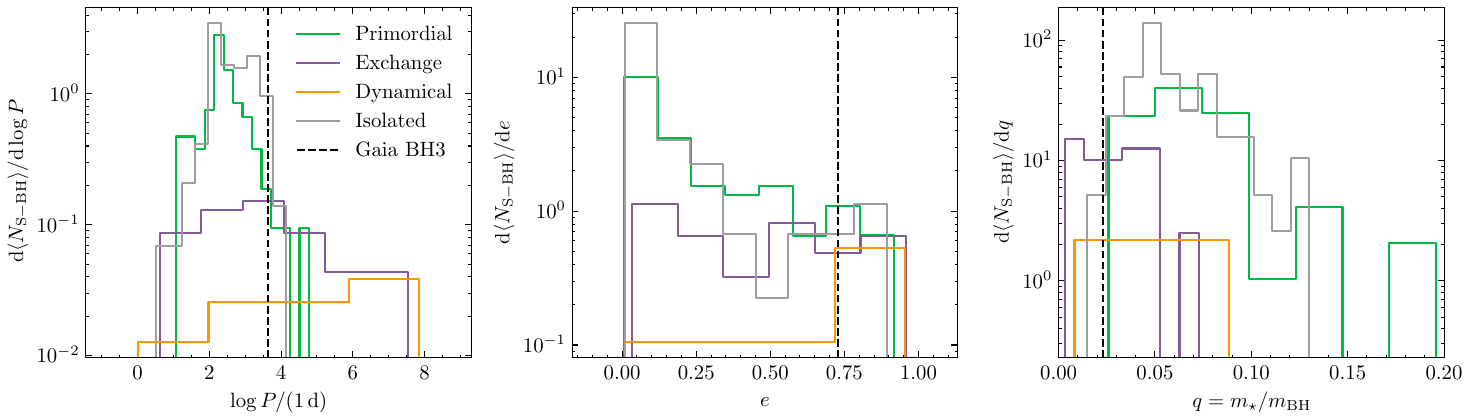}
    \caption{Probability distribution function of the number of \sbh binaries in our simulations as a function of the logarithm of the period (left), the eccentricity (middle), and the mass ratio (right). The vertical scale is normalised such that the area under the histogram equals the average number of \sbh binaries per cluster. In green, purple, orange; the \sbh formed in the primordial, exchange, and dynamical channel, respectively. In grey, the primordial binaries but evolved in isolation (isolated channel); in dashed black, the parameters of \gb.}
    \label{fig:properties}
\end{figure*}

\begin{figure*}[htbp]
    \centering
    \includegraphics[width=17cm]{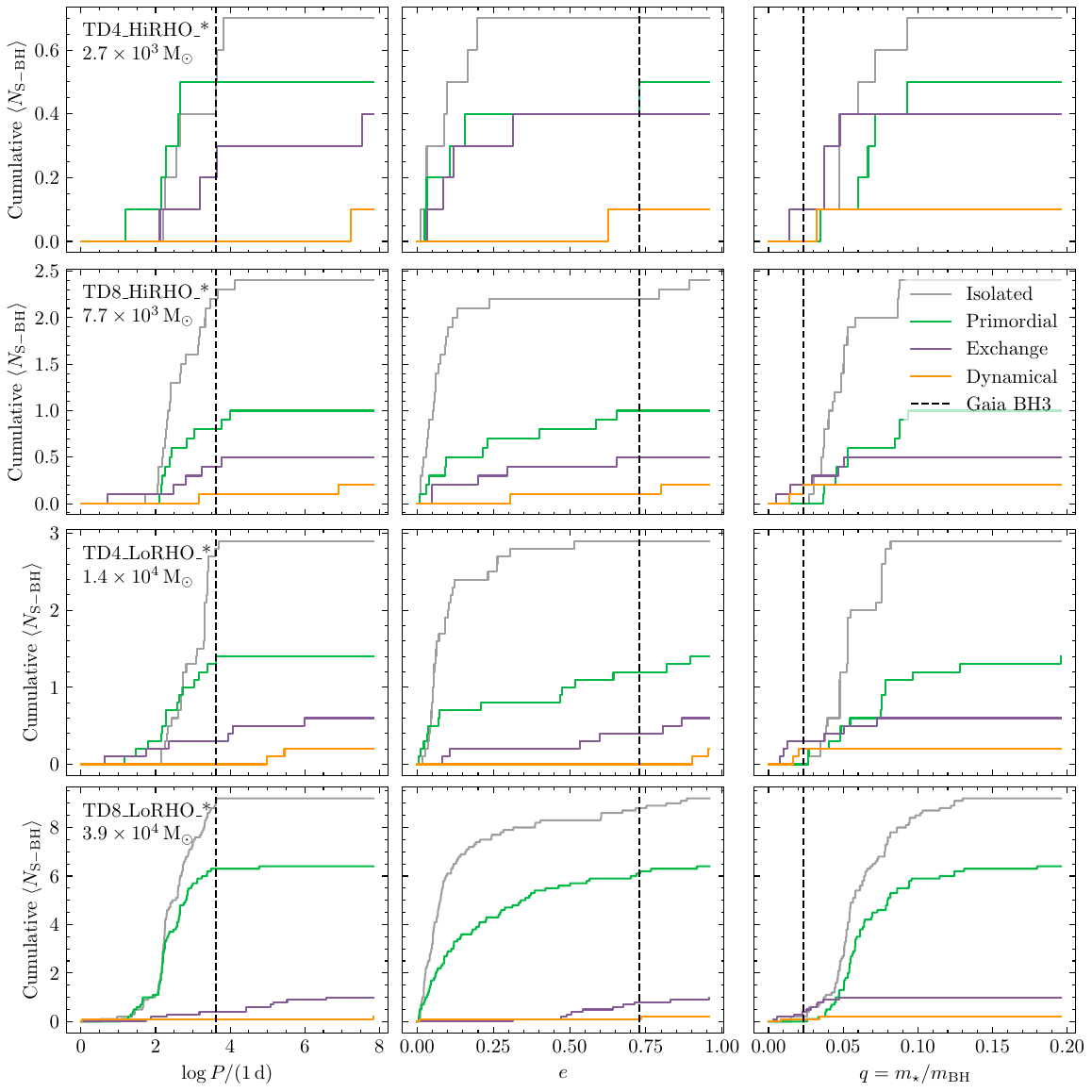}
    \caption{Cumulative distribution of the properties of \sbh binaries in our simulations as a function of the logarithm of the period (left), the eccentricity (middle), and the mass ratio (right), separated by the initial conditions of the model. The vertical scale is normalised such that the rightmost value is the mean number of \sbh binaries in each cluster simulation for each channel, obtained by averaging over $N_{runs}$. In green, purple, orange; the \sbh formed in the primordial, exchange, and dynamical channel, respectively. In grey, the primordial binaries but evolved in isolation; in dashed black, the parameters of \gb.}
    \label{fig:properties_model}
\end{figure*}

In this Section, we estimate the likelihood of the first three formation scenarios (primordial, exchange and dynamical).
We extract the data for these scenarios from the \sbh formed in our cluster simulations, defined as bound star-BH pairs present at the final snapshot (after $13\units{Gyr}$ of evolution), irrespective of their properties or location in the stream. For the primordial and exchange formation channels, we use the cluster models with primordial binaries (*\_WBIN), separating the \sbhs depending on whether they were paired up at the beginning of the simulation. For the dynamical formation scenario, we use our models without primordial binaries (*\_NOBIN). In Fig.~\ref{fig:properties}, we show the distribution of properties of the \sbh binaries in our models, separated by formation channel. In Fig.~\ref{fig:properties_model}, we show the same data but separated by $M_0$, $\rho_{h, 0}$, and $t_{dis}$, using cumulative histograms due to low-number statistics. In both these figures, we exclude the low-density models (TD0.5\_VLoRHO\_WBIN) since they can not reproduce the observed ED-2 stream data.

To discern the most likely formation scenario, we will compute the posteriors $P(\text{prim}\,|\,\thetaG)$, $P(\text{exc}\,|\,\thetaG)$, and $P(\text{dyn}\,|\,\thetaG)$, with $\thetaG=\{\log P, e, q\}$ the measured parameters of \gb. Each of these represents the probability that \gb formed in its respective scenario. Using Bayesian statistics, we have
\eq{P(\text{dyn}\,|\,\thetaG)=\frac{P(\thetaG\,|\,\text{dyn})P(\text{dyn})}{P(\thetaG)}}
with
\eq{\nonumber P(\thetaG)=\ & P(\thetaG\,|\,\text{prim})P(\text{prim})+P(\thetaG\,|\,\text{exc})P(\text{exc})\\ &+P(\thetaG\,|\,\text{dyn})P(\text{dyn})}
Here, $P(\text{prim})$, $P(\text{exc})$, and $P(\text{dyn})$ are the fraction of \sbhs formed in our simulations for each of the three channels, and are taken as priors. The priors $P(\text{prim})$, $P(\text{exc})$ are computed by dividing the total number of \sbh systems formed in the primordial and exchange channels (respectively) by the number of \textsc{*\_WBIN} clusters in our simulations. The prior $P(\text{dyn})$ is computed by dividing the total number of dynamical \sbh systems in the \textsc{*\_NOBIN} simulations by the number of these simulations. $P(\thetaG\,|\,\text{dyn})$ is the probability density function (PDF) of the parameters found in the dynamical channel, evaluated at $\thetaG$. The PDF is computed via a Kernel Density Estimation (KDE) using Gaussian kernels with a bandwidth estimated with Silverman's criterion. The posteriors for the other channels, $P(\text{prim}\,|\,\thetaG)$ and $P(\text{exc}\,|\,\thetaG)$, are computed using the same approach.

\begin{figure}
    \resizebox{\hsize}{!}{\includegraphics{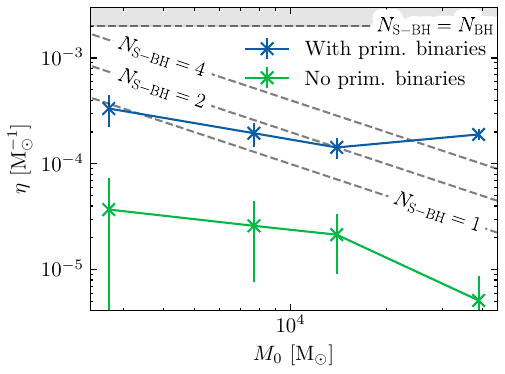}}
    \caption{Efficiency $\eta$, defined as the number of \sbh binaries $N_\sbh$ per unit of initial cluster mass, as a function of the initial cluster mass. In blue, models with a primordial binary population; in green, models without primordial binaries. The error bars represent the standard error of the mean (among multiple $N$-body models), assuming $N_\sbh$ behaves like a Poissonian variable.}
    \label{fig:efficiency}
\end{figure}

On average, there are significantly more \sbh systems in clusters with primordial binaries (see Table~\ref{table:Nsbh}). While we expect this to depend on the initial binary fraction, we find $P(\text{prim})\gg P(\text{dyn})$ and $P(\text{exc})\gg P(\text{dyn})$ (i.e. the rate of formation of dynamical binaries is very low). We show this in Fig.~\ref{fig:efficiency} and discuss it below. In our Bayesian analysis, the posteriors are $P(\text{exc}\,|\,\thetaG)=0.79$, $P(\text{prim}\,|\,\thetaG)=0.18$ and $P(\text{dyn}\,|\,\thetaG)=0.03$. These are the probabilities that \gb formed in each of the three channels, accounting for both the formation rates and the measured parameters (period, eccentricity, and mass ratio). Although the parameters of \gb are slightly better reproduced in a dynamical formation scenario \citep[as was found in][]{MarinPina2024}, the low efficiency of models without primordial binaries at producing \sbhs means that the dynamical formation scenario is disfavoured. Overall, our models show a preference for \gb being formed as an exchange binary. 

In our simulation with binaries, all the BH progenitor stars are originally paired in primordial binaries. This is due to our choice of initial conditions that forces a 100\% binary fraction for stars above $5\units{\msun}$. Within the exchange channel, we find that most of our \sbh binaries ($\sim 95\%$) form from a BH (or a binary containing a BH) that interacts with a low-mass stellar binary, exchanging the companion star and forming a dormant \sbh. 

\begin{table}[h]
\caption{Average number of \sbh binaries per cluster at the end of the simulation $\avg{N_\sbh}$, separated by cluster model} 
\label{table:Nsbh}
\centering 
\begin{tabular}{l c}
\hline\hline
Model name & $\avg{N_{\sbh}}$\\  
\hline   
TD8\_HiRHO\_WBIN &  1.5\\
TD8\_HiRHO\_NOBIN & 0.2 \\
TD8\_LoRHO\_WBIN & 7.4\\
TD8\_LoRHO\_NOBIN & 0.2\\
TD4\_HiRHO\_WBIN & 0.9 \\
TD4\_HiRHO\_NOBIN &0.1 \\
TD4\_LoRHO\_WBIN & 2.0 \\
TD4\_LoRHO\_NOBIN & 0.3\\
TD0.5\_VLoRHO\_WBIN & 1.2 \\
\hline
\end{tabular}
\end{table}

Furthermore, we can use the observed properties of \gb to constrain the initial properties of the ED-2 stream progenitor. Qualitatively, the data shown in Fig.~\ref{fig:properties_model} hint at a preference towards the massive ($M_0\geq 1.4\times 10^4\units{\msun}$) models with intermediate values of the initial density ($\rho_{h, 0}=240\units{\msun}\units{pc^{-3}}$); mostly due to the positive scaling of $\avg{N_\sbh}$ with initial cluster mass (Table~\ref{table:initial_params} and \ref{table:Nsbh}). This preference agrees with the results of the comparison with the ED-2 stream in Sect.~\ref{sec:stream}. While the most massive cluster models (TD8\_LoRHO\_WBIN) produce a handful of \sbh binaries ($\avg{N_\sbh}\simeq 7.4$, see Fig.~\ref{table:Nsbh}), most of these will not be detectable due to their distance to the Sun (because of the fanning of the stream) and \Gaia observational cuts.

In general, S-BH binaries are rare sources in clusters. In Fig.~\ref{fig:efficiency}, we show the efficiency, $\eta$, defined as the average number of S-BH per cluster (at 13 Gyr) per unit of initial cluster mass. Our simulations with primordial binaries have a few ($\sim1-4$, see Fig.~\ref{table:Nsbh}) of these binaries per cluster, whereas models without primordial binaries typically produce one or none. We conclude that we do not expect to detect other \sbhs in the ED-2 stream. Including primordial binaries increases efficiency by an order of magnitude; as mentioned above, we expect the formation rates to depend on the initial binary fraction. While our models predict that the efficiency is a slowly decreasing function of mass, we do not recover the scaling in \cite{MarinPina2024}, i.e. $\eta\propto M^{-1.2} r_v^{-0.42}$ with $r_v$ the initial virial radius. Instead, we find $\eta\propto M^{-0.5} r_v^{-0.1}$. This can be attributed to different choices of the initial conditions. For example, their models start with $\rho_{h, 0}\sim 10^{3-6}\units{\msun}\units{pc^{-3}}$ and much higher masses ($M_0 \sim 10^{5-6}\units{\msun}$). Compared to our models ($\rho_{h, 0}< 10^3\units{\msun}\units{pc^{-3}}$), they explore a different regime for the early dynamics of the cluster.

\subsection{Could \gb have formed in isolation?}
\label{ssec:iso}
In the previous Section, we showed that \gb likely formed as an exchange binary in the ED-2 progenitor cluster. We now test the effect that dynamical interactions in the cluster had on shaping the properties of primordial binaries. This is akin to asking how the properties of \gb are affected by dynamical interaction after it formed from binary evolution, as proposed in e.g.~\cite{Iorio2024, ElBadry2024, Mapelli2026}.

\begin{figure}
    \resizebox{\hsize}{!}{\includegraphics{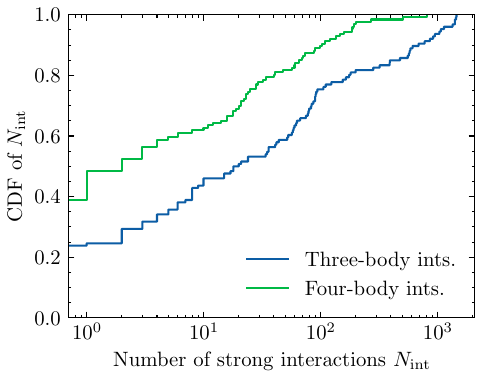}}
    \caption{Cumulative density function of the number of interactions undergone by the S-BH binaries present at the end of our simulations. In blue, binary-single (three-body) interactions; in green, binary-binary (four-body) interactions. The median value for three-body interactions is $N_{int}=18$, for four-body interactions it is $N_{int}=3$}
    \label{fig:Nint}
\end{figure}

\begin{figure}
    \resizebox{\hsize}{!}{\includegraphics{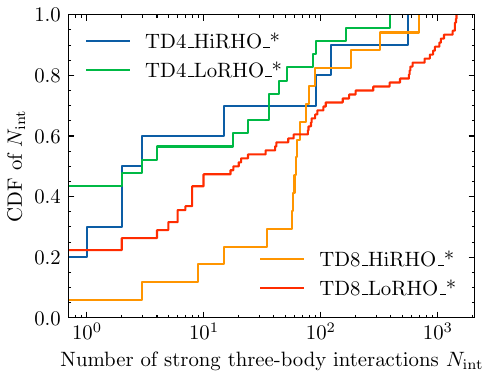}}
    \caption{Cumulative density function of the number of three-body interactions undergone by the \sbh binaries present at the end of our simulations, separated by the initial conditions of the cluster.}
    \label{fig:Nint_model}
\end{figure}

To quantify the effect of dynamical interactions on \sbh properties, we have extracted all binaries at the beginning of our cluster simulations and evolved them in isolation (i.e., unperturbed by cluster dynamics). We have used the same binary stellar evolution code and prescriptions as in the cluster simulations, so that all differences are strictly due to dynamical interactions. In Fig.~\ref{fig:properties}, we show the distributions of the properties of the \sbh systems formed in our simulations, separated by formation channels and including those binaries evolved without the effects of cluster dynamics (i.e., in isolation). We have also included the systems evolved in isolation in Fig.~\ref{fig:properties_model}.

We find that dynamics significantly modifies the \sbh binaries formed in the cluster models. In those with primordial binaries, at the end of the simulation, there are $\sim 22\%$ fewer S-BH than would be present if the binaries evolved in isolation. From Fig.~\ref{fig:properties}, disruptions affect long-period binaries the most. Dynamics leads to a marginal reduction of the number of \sbhs, but a significant change in their properties. 

Furthermore, in Fig.~\ref{fig:Nint}, we show the number of strong interactions, $N_{int}$, that each \sbh has undergone by the end of the simulations. We define strong interactions as those that undergo slow-down algorithmic regularisation (see \citealt{SDAR} and Sect.~\ref{sec:methods}). For a binary of mass $m_b$ with a perturber of mass $m_p$, this is implemented when the distance from the perturber to the closest member of the binary, $r_p$, satisfies the following condition
\eq{r_p<\lr{\frac{2m_p}{m_b\gamma_{min}}}^{1/3}R,}
where $R$ is the closest distance along the perturber's trajectory and $\gamma_{min}=1.5$ a free coefficient. For binary-single interactions, the median value is $N_{int}=19$, while for more chaotic binary-binary interactions, $N_{int}=3$. We can further this analysis by limiting our sample to \sbhs with a period similar to \gb. If we apply a conservative cut of $10^3<P/(1\units{d})<10^5$, we find a median of $N_{int}=28$ for three-body interactions and $N_{int}=1$ for four-body interactions. The difference is not significant, which can be interpreted as the maximum and minimum periods for \sbhs being the tails of the $N_{int}$ distribution. The number of strong interactions highlight the importance of dynamics and disfavour an isolated approach to studying the formation of \gb.

We now analyse the effect of the initial conditions of the cluster models on $N_{int}$ (Fig.~\ref{fig:Nint_model}). If we focus on three-body interactions, we can see a larger $N_{int}$ in the models with a larger dissolution time, and only a small correlation with density. Indeed, by instead analysing $N_{int}/t_{dis}$, we drive the models together, with the main difference between them being the initial density. Furthermore, the gap in the $N_{int}$ cumulative distribution caused by the difference in initial density is higher in models with a larger dissolution time. We note that this applies only to \sbh binaries that survive until the end of the simulations, and the cluster may have had \sbhs that underwent more interactions, but those binaries were disrupted or merged and thus are not present in our sample.

As an interesting addendum, strong interactions involving \sbh binaries may lead to observable transients such as micro-tidal disruption events \citep[$\mu$TDEs,][]{Perets2016, Rastello2025}.

\subsection{\gb as a triple system}
\label{sec:sbbh}
An alternative explanation for the nature of \gb is that its $33\units{\msun}$ BH is actually a BBH with a small SMA. In this section, we study the formation of star-BBH (S-BBH) systems in our ED-2 progenitor models.

If we assume that \gb is a S-BBH, there are some constraints on the possible ranges of SMA of the BBH orbit. First, if it is too small, the BBH will inspiral and merge due to gravitational wave (GW) emission. Our cluster simulations take this into account by implementing the orbit-averaged 2.5 post-Newtonian orbital shrinking \cite[Peters equation, ][]{Peters1964}. Thus, if we find a S-BBH at the present time in our simulations, it has survived for long enough without merging. On the other side, the SMA of the BBH can not be too large, as that would induce dynamical effects in the star that would have been detected by \Gaia astrometry. Future observations can provide a stronger upper bound on the SMA of a possible inner binary.

In our simulations, we find a very small amount of S-BBH systems, with \sbh binaries being $40$ times as common. Furthermore, none of the S-BBHs has an outer orbit compatible with the observed properties of \gb, mostly because they have much longer periods. Therefore, our models disfavour the hypothesis that the unseen companion in \gb is a BBH, in agreement with previous results \citep{Tanikawa2025}.

\section{S-BH formation environment}
\label{sec:formenv}
\begin{figure}
    \resizebox{\hsize}{!}{\includegraphics{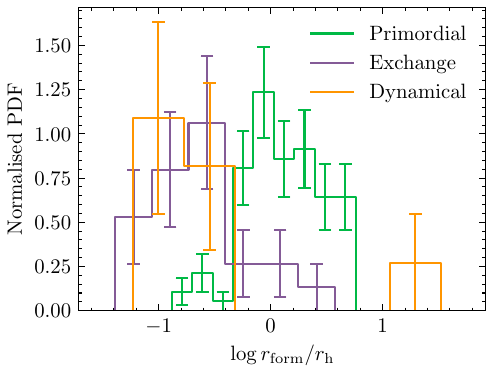}}
    \caption{Probability distribution function of the number of \sbh binaries in our simulations as a function of the logarithm of their formation radius (in units of the half-mass radius, $r_h$), separated by formation channels. The vertical scale is normalised such that the area under the histogram is one. The uncertainties of each bin are computed as $\sqrt{N_\sbh}$ within the bin.}
    \label{fig:rform}
\end{figure}

In this Section, we characterise the formation environment of the \sbh binaries in our simulations, i.e. their position, time, and dynamical hardness at the moment of formation. As in the previous Section, we consider only the S-BH present at the end of our simulations (i.e., after $13\units{Gyr}$ of evolution). First, we characterise the distribution of the radial position of S-BHs at formation, $r_{form}$. For each S-BH binary, we obtain the first snapshot they appear in and compute $r_{form}$ from their distance to the density centre of the cluster, defined as in \cite{CasertanoHut1985}. Results are shown in Fig.~\ref{fig:rform}. 

For primordial binaries, the $r_{form}$ distribution is simply the initial density profile of the cluster, with a median $r_{form}\sim r_h$, where $r_h$ is the half-mass radius of the cluster. For the exchange binaries, their formation happens at a rate (per unit volume) of $\Gamma \sim n_\BH n_{b} \Sigma v_{rel}$, with $n_\BH$, $n_b$ the local number density of stellar binaries and BHs, respectively; $\Sigma$ their relative cross section \citep[usually estimated from][]{Heggie1996}, and $v_{rel}$ their relative velocity. Around the central regions of the cluster, the density profile can be taken as roughly constant, so that $\Gamma$ is roughly independent of the radial position. Then, the number of formed exchange \sbh binaries is $N_\sbh \sim \Gamma r^3 \Delta t$, thus the distribution of $r_{form}$ is $\mathrm{d}N_\sbh/\mathrm{d}r_{form}\propto r^2$. However, at a certain radius, the density and interaction rates drop, thus producing the peak seen in Fig.~\ref{fig:rform}. A similar argument can be made for dynamical binaries.

\begin{figure}
    \resizebox{\hsize}{!}{\includegraphics{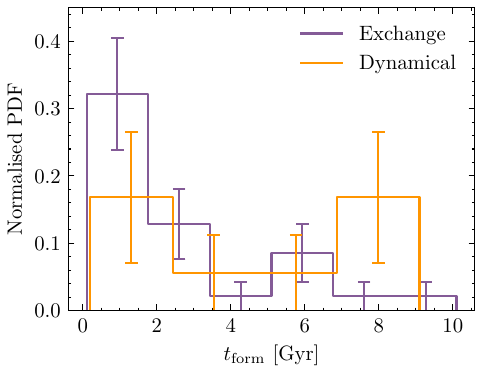}}
    \caption{Probability distribution function of the number of \sbh binaries in our simulations as a function of their formation time, separated by formation channels. The vertical scale is normalised such that the area under the histogram is one. Primordial S-BH binaries (not shown) form at $t=0$. The uncertainties of each bin are computed as $\sqrt{N_\sbh}$ within the bin.}
    \label{fig:tform}
\end{figure}

Next, we consider the formation time of S-BH binaries, $t_{form}$. We define it as the first snapshot in the simulations at which the star is paired in a binary with the BH (or its progenitor). For storage efficiency, we only save snapshots every $100\units{Myr}$; however, the error introduced in this approach is $\lesssim 1\%$ of our simulation time. The distribution of $t_{form}$ is shown in Fig.~\ref{fig:tform}. Dynamical binaries have a roughly constant, low formation rate; meanwhile, the formation of exchange binaries shows hints of peaking at earlier times.

An alternative formation channel suggests the possibility of assembling S-BH binaries in tidal tails \citep{Penarrubia2021}. In our set of simulations, we only find three S-BH binaries that become formally bound after cluster dissolution, with a SMA about $10^{2-3}$ times larger than that of \gb. While this is a promising formation pathway for wider binaries, we disfavour it as a formation channel of \gb.

Finally, we consider the hardness ratio, $\epsilon$, of S-BH binaries in our simulation, defined as the ratio of the (absolute) binary binding energy to the average kinetic energy of stars in the central region of the cluster, both evaluated at the moment of formation. Specifically, we compute it as $\epsilon=\abs{E_b}/\langle m \sigma^2 \rangle$, with $\abs{E_b}=G m_\star m_\BH/(2a)$, and $\langle m \sigma^2 \rangle$ computed within $r_h$. The value of $\epsilon$ defines a transition in the stability of binaries in clusters: those with $\epsilon \lesssim 1$ (soft binaries) tend to disrupt due to the interaction with other stars or BHs, while binaries with $\epsilon \gtrsim 1$ (hard binaries) tend to get more bound after an interaction. 

\begin{figure}
    \resizebox{\hsize}{!}{\includegraphics{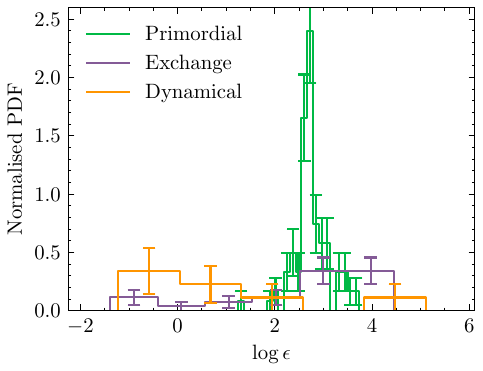}}
    \caption{Probability distribution function of the number of \sbh binaries in our simulations as a function of the logarithm of their hardness ratio $\epsilon$, separated by formation channels. The vertical scale is normalised such that the area under the histogram is one. The uncertainties of each bin are computed as $\sqrt{N_\sbh}$ within the bin.}
    \label{fig:hardness}
\end{figure}

In Fig.~\ref{fig:hardness}, we show $\epsilon$ for all S-BH binaries in our simulations. We find that $\sim90\%$ of the exchange binaries and $\sim 60\%$ of the dynamical binaries in our simulation form above the hard-soft boundary ($\epsilon \geq 1$). While soft binaries are likely formed at a much higher rate, they are easily ionised and thus do not survive to the end of the simulation. At the moment of cluster formation, the primordial binaries that will become primordial S-BHs are hard, with $\epsilon\sim 10^{2 - 3}$.

\section{Discussion}
\label{sec:discussion}
\subsection{Alternative binary evolution models}
\label{sec:discussion:bse}
The use of different assumptions for stellar and binary evolution (including the properties of primordial binaries) in our cluster simulations would result in different values for the parameters of \sbh systems formed in our models. In this Section, we argue that the key points in this paper are independent of the assumed evolutionary models and initial conditions.

First, our models show that the most likely origin for \gb is an exchange, which shows that the effect of dynamics is important for the formation of this binary. This discards the possibly unrealistic limit case with no primordial binaries. Furthermore, the comparison of the observable sources in our simulations (after detection cuts) to the stream data (Sect.~\ref{sec:stream}) shows that the cluster had an initially high density, meaning that any S-BH binary still in the stream after multiple Gyrs of evolution necessarily underwent multiple strong interactions (see Fig.~\ref{fig:Nint}). After these interactions, which include exchanges and resonances, the effect of the initial conditions for binaries is diluted.

Regarding the isolated scenario (where the binary evolved in the ED-2 progenitor unperturbed by dynamical interactions), we are not able to reproduce the mass ratio of \gb due to our initial choice for the limits in the mass ratio distribution. Choosing different limits and other binary evolution models, it has been shown that it is possible to form \gb-like systems \citep{ElBadry2024, Iorio2024, Mapelli2026}. However, our argument to disfavour an isolated origin of \gb is based on the ionisation and overall difference of the \sbh systems with the input initial conditions due to cluster dynamics. Since \gb formed in the ED-2 progenitor, it must have undergone multiple strong interactions whose effect can not be neglected.

\subsection{Uncertainties in the Galactic potential}
In our simulations, we have only considered a simplified model of the Galactic potential \citep[\textsc{mwpotential2014},][]{galpy}. This potential is defined by a bulge with a power-law density profile with an exponential cut-off, a disk, and a dark-matter halo. However, it is axisymmetric and does not consider time evolution. 

The use of that specific form of the potential is justified mainly by our ignorance about the time evolution of the true Milky Way potential, together with computational constraints. Including irregularities in the potential would lead to a heating of the ED-2 stream in our simulations, which implies a greater fanning of the stars along the direction transverse to the principal axis of the stream. This would play a role in Fig.~\ref{fig:stream}, which would bring the simulations closer to the observed data. Furthermore, this would lead to a shift of the distribution of $N_{vis}$ in Fig.~\ref{fig:Nvis_all} to lower values, which only strengthens our conclusions that the ED-2 stream formed from a cluster with a high initial density.

\section{Conclusions}
\label{sec:conclusions}
In this paper, we study different formation mechanisms of \textit{Gaia} BH3 in the progenitor cluster of the ED-2 stream. To do so, we use targeted $N$-body simulations of that cluster, including realistic single and binary stellar evolution models and the Galactic potential.

Regarding the stream formation, we show that the likely progenitor of the ED-2 stream is a now-dissolved star cluster, with an initial mass within the uncertainties presented in \cite{Balbinot2024}, i.e. $2\times 10^3\lesssim M/\msun \lesssim 5.2 \times 10^4$. The higher-mass limit is preferred, based on the number of visible sources in the stream. We show that this cluster dissolved after some Gyrs of evolution, and a scenario where it had a very low initial density is disfavoured, as that would lead to a stream much more sparse than what is found in the data.

Regarding the origin of \gb, we show that it likely formed from an exchange interaction between a BH and a primordial stellar binary of the ED-2 cluster, which had its properties significantly altered by subsequent interactions within the cluster environment. The probability that \gb formed as a dynamical binary from a previously-unbound BH and a single star is low. Furthermore, we show that the properties of primordial \sbh binaries in the ED-2 progenitor cluster are significantly modified by dynamical interactions, which challenges models that study the formation of \gb in isolation. We also consider the possibility that \gb is a triple system composed of an inner BBH with an outer star, but we confirm previous studies that show that this is an unlikely scenario.

Finally, we characterise the formation environment of S-BH binaries in our simulation, including their radial position, formation time, and binding energy. These results can be useful as hindsight for future studies targeting the formation of S-BH systems in clusters, including the potential for rapid population synthesis methods.

In summary, this paper shows that the most likely formation scenario for \gb is a BH formed as part of a primordial binary that exchanged its companion star, then underwent multiple strong dynamical interactions. We highlight the importance of cluster dynamics in assembling \gb, and disfavour a formation scenario where it evolved unperturbed by strong dynamical interactions. Our work suggests that future studies interpreting the population of dormant \sbh binaries of the next \textit{Gaia} Data Release can not neglect the effects of cluster dynamics.

\begin{acknowledgements}
    We thank Katie Breivik for insights regarding the isolated scenario. DMP thanks other participants in the KITP programme \textsc{stellarbh25} for useful discussions. DMP thanks Ugo di Carlo for his early involvement in the project. DMP thanks Ataru Tanikawa for insights regarding the S-BBH scenario. DMP acknowledges support from the Deutsche Forschungsgemeinschaft (DFG, German Research Foundation) through project number 546850815 (acronym: DoBlack) and under Germany's Excellence Strategy EXC 2181/1 - 390900948 (the Heidelberg STRUCTURES Excellence Cluster). This research was supported in part by grant no. NSF PHY-2309135 to the Kavli Institute for Theoretical Physics (KITP). DMP and MG acknowledge financial support from the grants PRE2020-091801, PID2024-155720NB-I00, CEX2024-001451-M, funded by MCIN/AEI/10.13039/501100011033 (State Agency for Research of the Spanish Ministry of Science and Innovation). SR acknowledges financial support from the Beatriu de Pinós postdoctoral fellowship program under the Ministry of Research and Universities of the Government of Catalonia (Grant Reference No. 2021 BP 00213). GI is supported by a fellowship grant from la Caixa Foundation (ID 100010434). The fellowship code is LCF/BQ/PI24/12040020. 
\end{acknowledgements}

\bibliographystyle{aa} 
\bibliography{bibliography} 

\begin{appendix}
    
\section{Dissolution time}
\label{sec:appendix:diss}
To find a combination of initial mass and half-mass density, $(M_0, \rho_{h,0})$, that results in a given dissolution time, $t_{dis}$,  we use the expressions for $t_{dis}$ of tidally limited clusters with BHs of \cite{GielesGnedin2023}

\eq{t_{dis}=10\units{Gyr} \frac{2/3}{y}\frac{30\units{\msun/Myr}}{\abs{\dot{M}_{ref}}}\frac{0.32\units{Myr^{-1}}}{\Omega_{tid}}\lr{\frac{M_i}{2\times 10^5 \units{\msun}}}^x.}
Here $\abs{\dot{M}_{ref}}$ is a reference mass loss rate and $x$ and $y$ are parameters that set how the mass-loss rate depends on initial and instantaneous mass, respectively (see equation~4 of \citealt{GielesGnedin2023}). These three parameters depend on $\rho_{h,0}$, relative to the tidal density, which depends on the strength of the tidal field, which is quantified by the parameter $\Omega_{tid}$.

For a single isothermal sphere,  $\Omega_{tid}=\sqrt{2}V_c/R_{eff}$, where $V_c=220\units{km/s}$ and $R_{eff}=R_p(1+e)$, with $R_p$ and $e$ the periapsis and eccentricity of the orbit of the cluster around the Galaxy, respectively. For our models, $R_p\simeq6.5\units{kpc}$ and $e\simeq0.6$, so $R_{eff}=10.5\units{kpc}$ and $\Omega_{tid}=0.03\units{Myr^{-1}}$. 

\cite{GielesGnedin2023} show that $x,y$ and $\abs{\dot{M}_{ref}}$ depend on $\rho_{h,0}$ relative to the tidal density, which can be captured by the parameter $\mathcal{R}_0\equiv \rho_{h, 0}/\rho_{h, f}$, where $\rho_{h, f}\simeq39 \Omega_{tid}^2/G\simeq8\,M_\odot\,{\rm pc}^{-3}$ is the half-mass density of a tidally filling cluster on the orbit of ED-2 (see section 2.1\footnote{We note that there is a typo in their expression for $\rho_{J,eff}$ on p. 5342 (left), where the 2 should be a 4.} in \citealt{GielesGnedin2023}).
For $t_{dis} = 4\,$Gyr and 8\,Gyr, we used at $\mathcal{R}_0 = 29$ and 95, and the corresponding values for $\abs{\dot{M}_{ref}}, x$ and $y$ from their table 1 to find $M_0$ and $\rho_{h,0}$. For $t_{dis} = 0.5\,$Gyr we used the lower $\mathcal{R}_0 = 3$, to ensure that $M_0$ is within the allowed range.

\end{appendix}
\end{document}